\newcommand{\rem}[1]{}
\documentclass[prb,twocolumn,showpacs,preprintnumbers,amsmath,amssymb,floatfix]{revtex4}
\usepackage{hyperref}
\usepackage{hypernat} 
\usepackage[dvips]{epsfig}

\begin{document}

\title{Tribology of the lubricant quantized-sliding state}

\author{Ivano Eligio Castelli$^1$, Rosario Capozza$^2$, Andrea Vanossi$^{3,2}$, Giuseppe E. Santoro$^{3,4}$, Nicola Manini$^{1,3}$, and Erio Tosatti$^{3,4}$}
\affiliation{
$^1$Dipartimento di Fisica and CNR-INFM, Universit\`a di Milano, Via
Celoria 16, 20133 Milano, Italy
}
\affiliation{
$^2$CNR-INFM National Research Center S3 and Department of Physics, \\
University of Modena and Reggio Emilia, Via Campi 213/A, 41100 Modena,
Italy
}
\affiliation{
$^3$International School for Advanced Studies (SISSA)
and CNR-INFM Democritos National Simulation Center, Via Beirut 2-4, I-34014
Trieste, Italy
}
\affiliation{
$^4$International Centre for Theoretical Physics (ICTP), P.O.Box 586,
I-34014 Trieste, Italy
}

\date{\today}

\begin{abstract}
In the framework of Langevin dynamics, we demonstrate clear evidence of the peculiar quantized sliding state, previously found in a simple 1D boundary lubricated model [Phys. Rev. Lett. {\bf 97}, 056101 (2006)], for a substantially less idealized 2D description of a confined multi-layer solid lubricant under shear.
This dynamical state, marked by a nontrivial ``quantized'' ratio of the averaged lubricant center-of-mass velocity to the externally imposed sliding speed, is recovered, and shown to be robust against the effects of thermal fluctuations, quenched disorder in the confining substrates, and over a wide range of loading forces. The lubricant softness, setting the width of the propagating solitonic structures, is found to play a major role in promoting in-registry commensurate regions beneficial to this quantized sliding. By evaluating the force instantaneously exerted on the top plate, we find that this quantized sliding represents a dynamical ``pinned'' state, characterized by significantly low values of the kinetic friction. While the quantized sliding occurs due to solitons being driven gently, the transition to ordinary unpinned sliding regimes can involve lubricant melting due to large shear-induced Joule heating, for example at large speed.

\end{abstract}


\maketitle

\section{Introduction}

When a confined lubricant is characterized by a large surface-to-volume ratio, the surface features of the confining walls may strongly affect its tribological properties.
If the width of the lubricant film is reduced to a few atomic layers, as in the boundary-sliding regime, the atoms in the film tend generally to be ordered into layers parallel to the bounding
substrates.\cite{Gao97,Tartaglino06}
Both numerical simulations and experiments generally conclude that, in such a strongly confined geometry, the film behaves like a solid, even at temperatures significantly higher than its bulk melting
temperature.\cite{Klein98}
In real operative conditions, moreover, the case of ``dry'' friction is quite exceptional.
A physical contact between two solids is generally mediated by so-called ``third bodies'', which act like a lubricant film.\cite{He99}
For crystals sliding on crystals, one may consider the moving contact as characterized schematically by three inherent length scales: the periods of the bottom and top substrates, and the period of the embedded lubricant structure.

The problem of boundary lubricated friction is fascinating both from the fundamental point of view and for applications.
Interesting dynamical behaviors, with possible tribologically important implications of an irregular distribution of the lubricant velocity in between the sliding surfaces, have recently been observed in numerical simulations, depending on the ``degree'' of geometrical incommensurability defining the moving
interface.\cite{Vanossi06,Manini07extended,Vanossi07PRL,Manini07PRE}
The prominent nontrivial feature of those simulation is an asymmetry in the relative sliding velocity of the intermediate lubricating sheet relative to the two substrates.
Strikingly, this velocity asymmetry takes an exactly quantized value which is uniquely determined by the incommensurability ratios and is insensitive to all other physical parameters of the model.
The occurrence of this surprising and robust regime of motion, giving rise to perfectly flat plateaus in the ratio of the time-averaged lubricant center-of-mass (CM) velocity to the externally imposed relative speed, was ascribed to the intrinsic topological nature of this quantized dynamics.
The phenomenon, investigated in detail in a rather idealized 1D
geometry,\cite{VanossiPRL,Santoro06,Cesaratto07,Vanossi07Hyst,Manini08Erice,Vanossi08TribInt}
was explained by the corrugation of a sliding confining wall rigidly
dragging the topological solitons (kinks or antikinks) that the embedded lubricant structure forms with the other substrate. Evidence of the existence of this peculiar regime of motion was then confirmed shortly after for a substantially less idealized 2D model of boundary lubrication, where atoms were allowed to move perpendicularly to the sliding direction and interacted via LJ potentials \cite{Castelli08Lyon}. The solitonic transverse corrugations of the lubricant, propagating inside the film from bottom to top, were observed to favor the kink (or antikink) tendency to pin to the top-layer spatial periodicity, thus strengthening the quantization mechanism.

In this work, we investigate the tribological signature of the quantized-velocity state for such a 2D multi-layer solid lubricant under shear. Its robustness against thermal effects (implemented by means of a standard Langevin approach), quenched disorder in the confining substrates, and over a wide range of loading forces is analyzed in detail. The lubricant softness, setting the width of the propagating solitonic structures, together with the soliton coverage number, is considered in promoting in-registry commensurate regions beneficial to the quantized sliding. Hysteretic termination and kinetic friction are properly evaluated to highlight tribological differences with respect to ordinary unpinned sliding regimes.

The paper is organized as follows. In Sec.~\ref{model:sec}, we summarize the main features of the 2D-implemented tribological confined model, together with the numerical procedure adopted to solve the system of coupled equations describing the dissipative particle dynamics and top plate motion at finite temperatures.
Section ~\ref{plateaudynamics:sec} is devoted to prove the robustness of the quantized dynamical regime against many physical parameters of the model.
In Sec.~\ref{softness:sec} the relative lubricant-lubricant and lubricant-substrate interactions are varied to study the role of the confined film softness.
By evaluating the force instantaneously exerted on the advancing plate in Sec.~\ref{friction:sec}, we characterize the tribological features of the quantized dynamics and show that it corresponds to significantly low kinetic friction force.
The transition towards ordinary sliding regimes can involve lubricant melting due to large shear-induced Joule heating, leading to a
higher-friction sliding state.
As detailed in Sec.~\ref{softnessintermittent:sec}, within the framework of our rigid, constant-speed driving, we observe an intermittent stick-slip dynamics, associated to tiny amplitude fluctuations, in the kinetic friction force experienced by the top substrate while in the quantized-velocity state.
Discussions and conclusions are given in Sec.~\ref{conclusions:sec}.

\section{The 2D confined lubricated model}\label{model:sec}

\begin{figure}
\centerline{
\epsfig{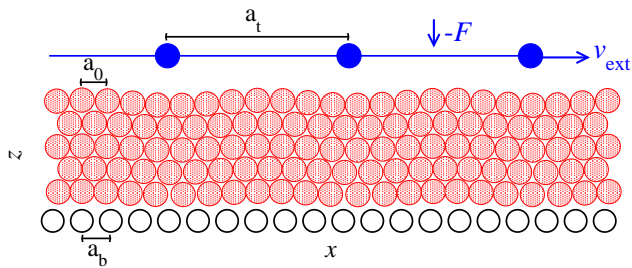}
}
\caption{\label{model:fig} (color online).
A sketch of the model with the rigid top (solid circles) and bottom (open)
crystalline sliders (of lattice spacing $a_{\rm t}$ and $a_{\rm b}$
respectively), the former moving at externally imposed $x$-velocity $v_{\rm
  ext}$.
One or more solid
lubricant layers (shadowed) of rest equilibrium spacing $a_{\rm
0}$ are confined in between.
}
\end{figure}

Our system is composed of two rigid walls made of equally-spaced atoms and
$N_{\rm l}$ identical lubricant atoms confined in between, organized in
$N_{\rm layer}$ layers, see Fig.~\ref{model:fig} where $N_{\rm layer}=5$.
The lubricant atoms move under the action of pairwise Lennard-Jones (LJ)
potentials
\begin{equation}\label{LJpotential}
\Phi_a(r)=\epsilon_a
\left[\left(\frac{\sigma_a}{r}\right)^{12}
-2\left(\frac{\sigma_a}{r}\right)^6\right]
\end{equation}
describing the reciprocal interactions.
The cutoff radius is set at $r=r_{\rm c}=
2.49\,\sigma_a$, where $\Phi_a\left(r_{\rm c} \right) \simeq -8.4\,\cdot
10^{-3}\,\epsilon_a$.

For the two substrates and the lubricant we assume three different kinds of
atoms, and characterize their mutual interactions ($\Phi_{\rm bl}$,
$\Phi_{\rm ll}$ and $\Phi_{\rm tl}$ refer to potential energies for the
bottom-lubricant, lubricant-lubricant, and top-lubricant interactions,
respectively) with the following LJ radii $\sigma_a$
\begin{equation}\label{sigma}
\sigma_{\rm tl}=a_{\rm t}\,,\qquad
\sigma_{\rm bl}=a_{\rm b}\,,\quad
{\rm and} \
\sigma_{\rm ll}=a_{\rm 0} \,,
\end{equation}
which, for simplicity, are set to coincide with the fixed spacings $a_{\rm
t}$ and $a_{\rm b}$ between neighboring substrate atoms, and the average
$x$-separation $a_{\rm 0}$ of two neighboring lubricant atoms,
respectively.
This restriction is only a matter of convenience, and is not essential to
the physics we are describing.
These three different periodicities $a_{\rm t}$, $a_{\rm 0}$ and $a_{\rm
b}$ define two independent ratios:
\begin{equation}\label{ratiotb}
\lambda_{\rm t}=\frac{a_{\rm t}}{a_{\rm 0}}\,,\qquad
\lambda_{\rm b}=\frac{a_{\rm b}}{a_{\rm 0}}\,,
\end{equation}
the latter of which we take closer to unity, $\max(\lambda_{\rm
b},\lambda_{\rm b}^{-1})<\lambda_{\rm t}$, so that the lubricant is less
off-register to the bottom substrate than to the top.

For simplicity, we fix the same LJ interaction energy $\epsilon_{\rm tl}
=\epsilon_{\rm ll} =\epsilon_{\rm bl} =\epsilon$ for all pairwise coupling
terms and the same mass $m$ of all particles.
We take $\epsilon$, $a_{\rm 0}$, and $m$ as energy, length, and mass units.
This choice defines a set of ``natural'' model units for all physical
quantities: for instance velocities are measured in units of
$\epsilon^{1/2} \, m^{-1/2}$.
In the following, all mechanical quantities are expressed implicitly in the
respective model units.

The interaction with the other lubricant and sliders' particles
produces a total force acting on the $j$-th lubricant particle,
of coordinate $\vec r_j$, given by
\begin{eqnarray}\label{Fj}
\vec F_j&=& -\frac{\partial}{\partial \vec r_j}\Big[
\sum_{i=1}^{N_{\rm t}}
\Phi_{\rm tl}\!\left(|\vec r_j-\vec r_{{\rm t}\,i}|\right) +
\\ \nonumber
&&
+ \sum_{j'=1 \atop j'\ne j}^{N_{\rm l}}
\Phi_{\rm ll}\!\left(|\vec r_j-\vec r_{j'}|\right)
+\sum_{i=1}^{N_{\rm b}}
\Phi_{\rm bl}\!\left(|\vec r_j-\vec r_{{\rm b}\,i}|\right)
\Big]
,
\end{eqnarray}
where $\vec r_{{\rm t}\,i}$ and $\vec r_{{\rm b}\,i}$ are the positions of
the $N_{\rm t}$ top and $N_{\rm b}$ bottom atoms.
By convention, we select the frame of reference where the bottom layer is
static, with the atoms located at
\begin{equation}\label{xzbottom}
r_{{\rm b}\,i\,x}(t)= i\,a_{\rm b}
\,,\qquad
r_{{\rm b}\,i\,z}(t)= 0
\,.
\end{equation}
The top slider is rigidly driven at a constant horizontal velocity
$v_{\rm ext}$, and can also move vertically (its inertia equals
the total mass $N_{\rm
  t}m$ of its atoms) under the action of the external loading force $F$
applied to each particle in that layer and that due to the interaction with
the particles in the lubricant layer:
\begin{equation}\label{xztop}
r_{{\rm t}\,i\,x}(t)= i\,a_{\rm t}+v_{\rm ext}\,t
\,,\qquad
r_{{\rm t}\,i\,z}(t)= r_{{\rm t}\,z}(t)
\,,
\end{equation}
where the equation governing $r_{{\rm t}\,z}$ is
\begin{eqnarray}\label{zztop}
N_{\rm t}m \,\ddot r_{{\rm t}\,z}
&=&\!-\! \sum_{i'=1}^{N_{\rm t}}\sum_{j=1}^{N_{\rm l}}
\frac{\partial \Phi_{\rm tl}}{\partial r_{{\rm t}\,i'\,z}}
\!\left(|\vec r_{{\rm t}\,i'}-\vec r_j|\right)-\!N_{\rm t}F
\,.
\end{eqnarray}

To remove the Joule heat, and
control the lubricant temperature in this driven system, rather than a
Nos\'e-Hoover thermostat \cite{nose-hoover,Martyna92} as in
Ref.~\onlinecite{Castelli08Lyon}, we use a standard implementation of the
Langevin dynamics,\cite{Gardiner} with the addition of a damping term plus
a Gaussian random force $\vec f_j(t)$ to the Newton equations for the
lubricant particles.
The damping force includes the contributions of the energy dissipation into
both individual substrates:
\begin{equation}\label{ffriction}
\begin{array}{rcl}
\vec f_{{\rm damp}\ i} =
 -\eta \, {\dot{\vec r}}_i -\eta \, ({\dot{\vec r}}_i -\dot{{\vec r}}_{\rm t})
\,.
\end{array}
\end{equation}
Taking into account this double contribution to the $\eta$-dissipation,
the Gaussian null-average random forces satisfy the relation
\begin{equation}\label{2fd}
\langle
f_{j\,x}(t)\,f_{j\,x}(t')\rangle = 4 \eta k_{\rm B} T \, \delta(t-t')
\end{equation}
and similarly for the $\hat z$ components, such that in a static configuration
($v_{\rm ext}=0$) the Langevin thermostat leads to a steady state
characterized by standard Boltzmann equilibrium
\begin{equation}\label{equilibrium}
\langle E_k\rangle=2 \, N_{\rm l} \, \frac 12\, k_{\rm B} T
\,,
\end{equation}
with the total kinetic energy of the lubricant $E_k=\frac 12 \sum_i m {\dot
  {\vec r}}_i^{\,2}$.
This algorithm represents a simple but numerically stable and effective
phenomenological approach to describe energy dissipation into the
substrates occurring through the excitation of phonons and (in the case of
metals) of electron-hole pairs.
As we shall see below, the Langevin thermostat controls and compensates
reasonably well the expected Joule heating of the substrate, even in the
sliding regime ($v_{\rm ext}\neq 0$).
We adopt a fairly small value $\eta=0.1$ leading to underdamped lubricant
dynamics.
To guarantee a detailed force balance (Newton's third law), we add the
following force term
\begin{equation}\label{fextratop}
\eta \, \sum_i^{N_{\rm l}}({\dot{\vec r}}_i-\dot{{\vec r}}_{\rm t})
=
\eta \, N_{\rm l} \,(\vec v_{\rm cm}-\dot{{\vec r}}_{\rm t})
\end{equation}
to the equations for the motion of the top layer.
While the $\hat z$-component of this additional term has a real influence
on the motion governed by Eq.~\eqref{zztop}, of course its $\hat
x$-component only affects the external force $F_k$ required to maintain the
velocity component $\dot r_{{\rm t}\,x}$ constant and equal to $v_{\rm
  ext}$.

We integrate the ensuing equations of motion within a $x$-periodic box of
size $L=N_{\rm l} \, a_0$, by means of a standard fourth-order Runge-Kutta
method.\cite{NumericalRecipes}
We usually start off the dynamics of a single lubricant layer from
equally-spaced lubricant particles at height $r_{i\,z}=a_{\rm b}$ and with
the top layer at height $r_{{\rm t}\,z}= a_{\rm b}+a_{\rm t}$, but we
consider also different initial conditions.
For several layers, we initialize the system with lubricant particles
at perfect triangular lattice sites, and the top slider
correspondingly raised.
After an initial transient, sometimes extending for several hundred time
units, the sliding system reaches its dynamical steady state.
In the numerical simulations, an adiabatic variation of the external
driving velocity (a controllable parameter in tribological experiments) is
considered and realized by changing $v_{\rm ext}$ in small steps $\delta
v=0.1$, letting the system evolve at each step for a time long enough for
all transient stresses to relax, in practice 1500 time units.
In the calculations presented below, we compute accurate time-averages of
the physical quantities of interest by averaging over a simulation time of
3000 time units or more after the transient is over.
At higher temperature, fluctuations of all physical quantities around their
mean values increase, sometimes requiring even longer simulation times to
obtain well-converged averages.
To estimate error bars of all average quantities, we split the whole
trajectory into 30 pieces of equal duration, and then evaluate the standard
deviation of the averages carried out over each individual interval.

\section{The robustness of the plateau dynamics }\label{plateaudynamics:sec}

Here we prove the robustness of the quantized velocity plateau dynamics
against thermal fluctuations, changes in vertical load and the effect of
quenched disorder in the confining substrates.
In our simulations, we take into account complete layers, realizing an
essentially crystalline lubricant configuration at the given temperature,
generally kept below its melting temperature.
To investigate the dragging of kinks and the ensuing exact
velocity-quantization phenomenon, we evaluate the ratio $w=\overline{ v_{{\rm cm}\,x}} /v_{\rm ext}$ of the time-averaged lubricant
CM sliding velocity to the externally imposed sliding
speed $v_{\rm ext}$ stays pinned to an exact geometrically determined plateau value, while $v_{\rm ext}$ itself or temperature $T$ or even
relative substrate corrugation amplitudes are made vary over wide ranges.
In detail, the plateau velocity ratio
\begin{equation}\label{wplat}
w_{\rm plat}=
\frac{\overline{ v_{{\rm cm}\,x}}}{v_{\rm ext}}=
\frac{\frac 1{a_0} - \frac 1{a_{\rm b}} }{\frac 1{a_0}}=
\frac{\lambda_{\rm b} -1 } {\lambda_{\rm b}}=
1-\frac{1}{\lambda_{\rm b}}
\end{equation}
is a function uniquely determined by the excess linear density of
lubricant atoms with respect to that of the bottom substrate, thus
of the length ratio $\lambda_{\rm b}$, Eq.~(\ref{ratiotb}).
Despite not affecting the $w_{\rm plat}$ value, the top length ratio
$\lambda_{\rm t}$ also plays an important role, since it sets the kink
coverage $\Theta= N_{\rm kink}/N_{\rm t}= \left(1-\lambda_{\rm
  b}^{-1}\right) \lambda_{\rm t}$: assuming that the 1D mapping to the FK
model sketched in Ref.~\onlinecite{Vanossi07PRL} makes sense also in the present richer 2D geometry, the coverage ratio $\Theta$ should affect the pinning strength of kinks to the top corrugation, thus the robustness of the velocity plateau, as we will show in Sec.~\ref{doublelayers:sec} below.

As a convenient system for practical calculations
we consider a bottom substrate made of $25$ particles,
and $29$ lubricant particles in each layer, i.e. $\lambda_{\rm b}
=29/25=1.16$, which produces $4$ kinks every $29$ lubricant particles
in each layer. This value of $\lambda_{\rm b}$ is not to be considered
in any way special: we found perfect plateau sliding for many
other values of $\lambda_{\rm b}$. Moreover we also investigated the
plateau dynamics for an anti-kink configuration
$\lambda_{\rm b}=21/25=0.84$.

\begin{figure}
\centerline{
\epsfig{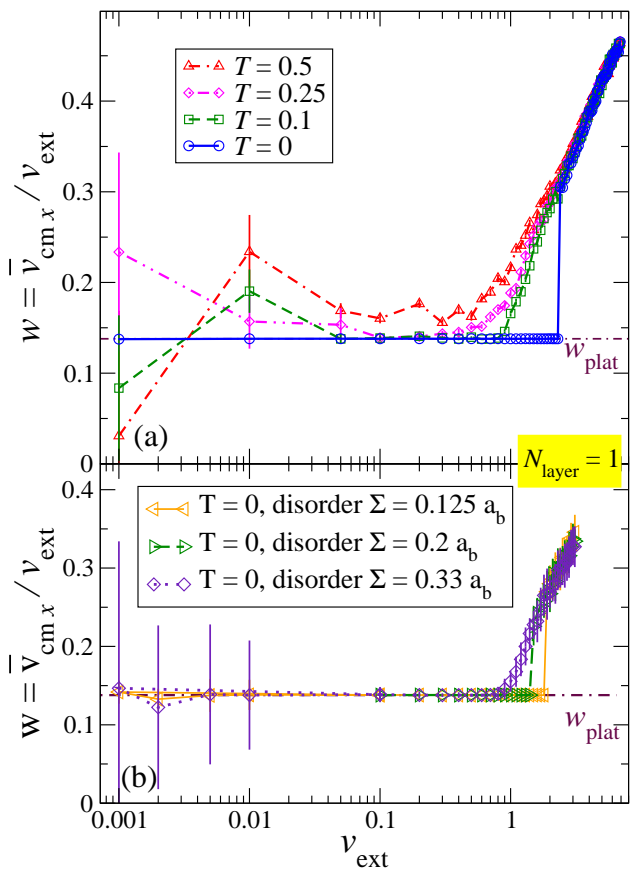} }
\caption{\label{vextmonotheta1:fig} (color online).
The time-averaged velocity ratio $w=\overline{v_{{\rm cm}\,x}} /v_{\rm ext}$ of a single lubricant layer as a function of the adiabatically increased top-slider velocity $v_{\rm ext}$ for different temperatures of the Langevin thermostat (panel (a)) and for distinct degrees of quenched disorder in the bottom substrate (panel (b)); atomic random displacements are taken in a uniform distribution in the interval $[-\Sigma/2,\Sigma/2]$ horizontally and
$[-\Sigma/4,\Sigma/4]$ vertically away from the ideal positions of
Eq.~\eqref{xzbottom}.
All simulations are carried out with a model composed by $4$, $29$ and $25$ atoms in the top lubricant and bottom layers respectively, with an applied load $F=25$.
The plateau velocity ratio (dot-dashed line) is $w_{\rm plat}
=\frac{4}{29}\simeq 0.138$, Eq.~(\ref{wplat}).
}
\end{figure}

Figure \ref{vextmonotheta1:fig}(a) reports the time-averaged horizontal
velocity $\overline{v_{{\rm cm}\,x}}$ of the single-layer lubricant CM, as a
function of the velocity $v_{\rm ext}$ of a fully commensurate top layer
($\Theta=1$) for several temperatures set by the Langevin thermostat.
The velocity ratio $w=\overline{v_{{\rm cm}\,x}}/v_{\rm ext}$ is generally
a nontrivial function of $v_{\rm ext}$, that for low temperature (here
$T=0$ and $T=0.1$) displays a wide flat plateau followed by a regime of
continuous evolution.
The plateau extends over a wide range of external driving velocities, up to
a critical depinning velocity $v_{\rm crit}$, whose precise value is
obtained by ramping $v_{\rm ext}$ ``adiabatically''.
Beyond $v_{\rm crit}$, the lubricant leaves the plateau speed and, at finite
temperature, moves up toward the symmetrical speed $w=0.5$, as
dictated by the thermostat dissipation.
At zero temperature, the plateau state is abandoned with a discontinuous
jump of $w$.
The small-$v_{\rm ext}$ side of the plateau is difficult to address by
numerical simulations, since the relative uncertainty in the determination
of $w$ increases due to $v_{\rm ext}$-independent thermal fluctuations of
$v_{{\rm cm}\,x}(t)$.
By running longer averaging simulations at low $v_{\rm ext}$, we mitigate
the error bars over $w$, and the data are consistent with a plateau
dynamics extending all the way down to the static limit $v_{\rm
  ext}\rightarrow 0$, like in the 1D model.\cite{Manini07PRE}
In particular, within the plateau, the $T=0$ calculation shows perfect
matching of the mean lubricant velocity to the geometric ratio $w_{\rm
  plat}$ of Eq.~(\ref{wplat}).
By increasing the temperature, $w$ deviates more and more from the perfect
plateau value.
At the highest temperature considered, $k_BT=0.5\,\epsilon$, near melting
of the LJ crystal at $k_BT_m \sim 0.7\,\epsilon$,\cite{Ranganathan92} and even at the smaller
$k_BT=0.25\,\epsilon$ no strict plateau is observed in the simulations.
Note however that for $v_{\rm ext}\lesssim 0.5$ the plateau attractor tends
to affect the dynamics even in the presence of large thermal fluctuations.
We have checked that finite-size scaling shows essentially no size effect
on the plateau, and in particular on its boundary edge $v_{\rm crit}$.

The dynamical pinning
mechanism of the present 2D model is similar to that discussed for the
simpler 1D model.\cite{Vanossi06,Manini07PRE}
The bottom layer produces a corrugated potential energy which, with its
near-matching periodicity, is responsible for the creation of kinks,
generally consisting of weak local compressions of the lubricant chain,
with two lubricant atoms trapped in the attractive region between two
bottom-substrate atoms.
In the quantized-velocity state, these kinks become pinned to the upper
slider (whose potential varies slowly on the lubricant interparticle scale)
and are therefore dragged along.

We probe the robustness of the quantized plateau state even in the
presence of an ``irregular'' (i.e., quenched-disordered) bottom substrate, by running simulations with its atoms
randomly displaced, horizontally and vertically, away from the regular
positions, Eq.~\eqref{xzbottom}, by a significant fraction of the lattice
equilibrium spacing $a_{\rm b}$.
Calculations still display the perfect plateau velocity, which however
terminates at a lower $v_{\rm crit}$.
Figure~\ref{vextmonotheta1:fig}b shows the results of one specific
realization of the quenched disorder of 3 different amplitudes $\Sigma$.
By averaging over different realizations of disorder with the same
amplitude $\Sigma=0.2\, a_{\rm b}$, we obtain, for example, $v_{\rm crit} =
1.47\pm0.27$, compared to the value $v_{\rm crit} \simeq 2.3$ obtained for
the case a perfectly crystalline bottom wall at $T=0$.

\begin{figure}
\centerline{
\epsfig{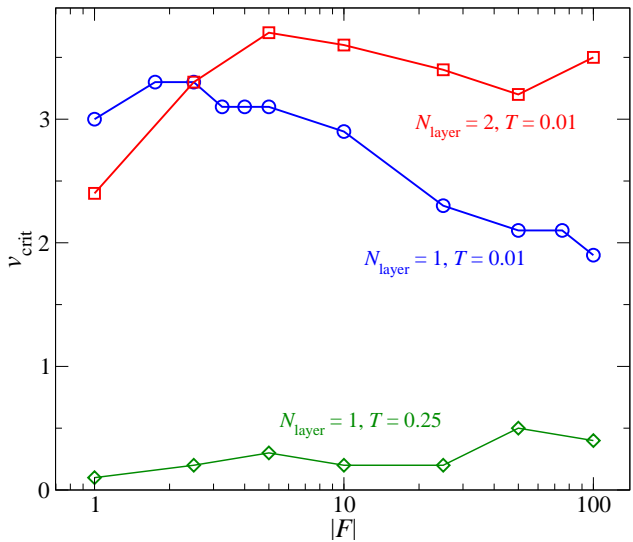} }
\caption{\label{Fmonotheta1:fig} (color online).
The critical depinning velocity for the breakdown of the quantized plateau
as a function of the applied load per top-layer atom $F$.
For the 1-layer curves, the same geometry as in
Fig.~\ref{vextmonotheta1:fig}, temperatures $k_{\rm B} T=0.01$ (circles)
and 0.25 (diamonds) are considered.
The 2-layer curve (squares) is computed with a doubled number of lubricant
atoms (58 rather than 29) confined between the same substrates in the same
$x$-range.
}
\end{figure}

Our overall evidence is
that, in the 2D system as much as in the more idealized
1D model, the phenomenon of velocity quantization is stable, reproducible
and ubiquitous, 
and not
the result of some careful tuning of
parameters.
In particular, Fig.~\ref{Fmonotheta1:fig} shows that the depinning velocity
marking the end of the quantized plateau is a smooth function of the load
$F$ applied to the top layer over two orders of magnitude.
Calculations suggest that a moderate load near unity is beneficial to the
quantized sliding, at least at the low temperature $k_{\rm B}T =
0.01\epsilon$.
This is the result of a
competition between the beneficial role
of load that limits thermally induced ``slips'' of the kinks, and
the detrimental effect of limiting the vertical lubricant
movements at high loads.
Indeed, at higher temperature, where thermal fluctuations are more active
in destabilizing the quantized state (thus yielding a smaller $v_{\rm
  crit}$), calculations show a general slow increase of $v_{\rm crit}$ with
$F$, favored by the suppression of thermally induced ``slips'' promoted by
the stronger load-induced confinement.

Figure~\ref{Fmonotheta1:fig} shows that in the large-load low-temperature
regime, the critical velocity for $N_{\rm layer}=2$ is even larger than for
$N_{\rm layer}=1$.
In this plateau regime, we identify ``horizontal'' kinks in the lubricant
layer adjacent to the bottom potential, while the upper lubricant layer
shows weaker $x$-spacing modulations.

\section{Soliton coverage and lubricant softness}
\label{softness:sec}
\label{doublelayers:sec}
The perfect matching of the number of kinks to the number of
top-atoms $\Theta=N_{\rm kink}/N_{\rm t}=1$, as considered in all
previous figures, is an especially favorable
circumstance
for kink dragging, thus
for the quantized-plateau phenomenon, but not
one
to be expected to
occur easily in actual lubricated sliding.
It is therefore important to examine situations where this
kind of full commensuration is 
absent.
\begin{figure}
\centerline{
\epsfig{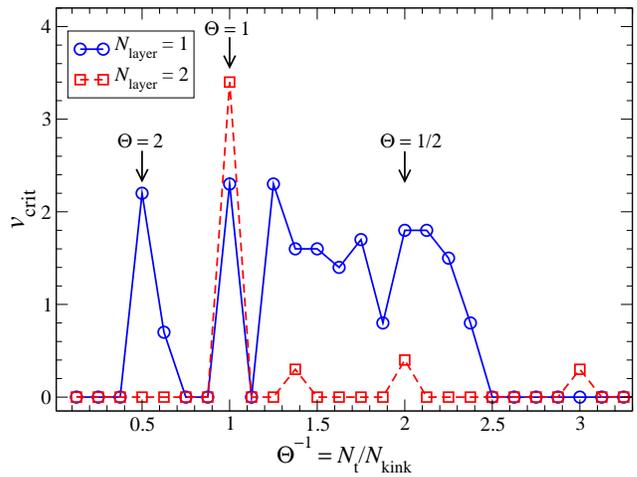}
} \caption{\label{vcrit:fig} (color online).
Variation of the plateau boundary velocity $v_{\rm crit}$ as a function
of the kink coverage $\Theta^{-1}=N_{\rm t}/N_{\rm kink}$ for a lubricant
monolayer and bilayer.
Calculations show local maxima of $v_{\rm crit}$ for commensurate
values of $\Theta$ for both $N_{\rm layer}=1$ and $2$; except at kink
coverage $\Theta=1$ $v_{\rm crit}$ is generally larger for $N_{\rm
  layer}=1$ than $N_{\rm layer}=2$.
Simulations are carried out with $F=25$, $T=0.01$,
and $\lambda_{\rm b}=\frac{29}{25}$.
}
\end{figure}

In the following we study the behavior of the depinning point $v_{\rm
crit}$ as a function of the commensuration ratio $\Theta$.
In particular we keep $\lambda_{\rm b}$ fixed (i.e. the density of solitons),
and vary the number of surface atoms in the top substrate.
In Fig.~\ref{vcrit:fig} the depinning velocity $v_{\rm crit}$, evaluated by
adiabatically increasing of $v_{\rm ext}$, is reported as a function of the
commensuration ratio $\Theta^{-1}=N_{\rm t}/N_{\rm kink}$.
While the dependence on kink coverage is rather erratic, we can recognize
that $v_{\rm crit}$ tends to peak at or near integer values of $\Theta$.
In contrast, a fractional $\Theta$ displaced from commensurate
regimes produces a weakening or even the loss of the quantized
plateau, represented by $v_{\rm crit}\simeq 0$.
Figure~\ref{vcrit:fig} shows that this drop in $v_{\rm crit}$ is
especially sharp for $N_{\rm layer}=2$, for which many
incommensurate coverages show no quantized plateau, even at the
smallest $v_{\rm ext}$ accessible within practical simulation
times.
We conclude
that the occurrence of quantized lubricant velocity plateaus,
although comparably robust and widespread, may not be trivial to observe in
practice: strongly incommensurate coverages and/or thick lubricant films
could prevent its manifestation.

\begin{figure}
\centerline{
\epsfig{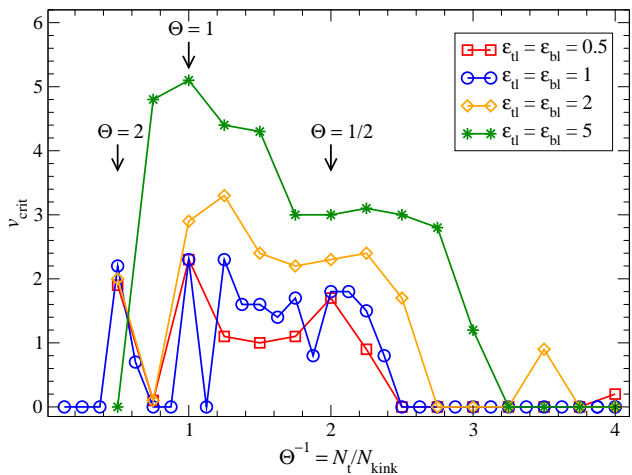}
} \caption{\label{eps_tp_bp:fig} (color online).
Variation of the plateau boundary velocity $v_{\rm crit}$ as a function
of the kink coverage $\Theta^{-1}=N_{\rm t}/N_{\rm kink}$ for a lubricant
mono-layer showing different interaction energies with the substrates.
Simulations are carried out with the same conditions as in
Fig.~\ref{vcrit:fig}, but with varied $\epsilon_{\rm tl} =\epsilon_{\rm
  bl}$, (but unchanged interparticle interaction $\epsilon_{\rm
  ll}=\epsilon$).
}
\end{figure}

\begin{figure}
\centerline{
\epsfig{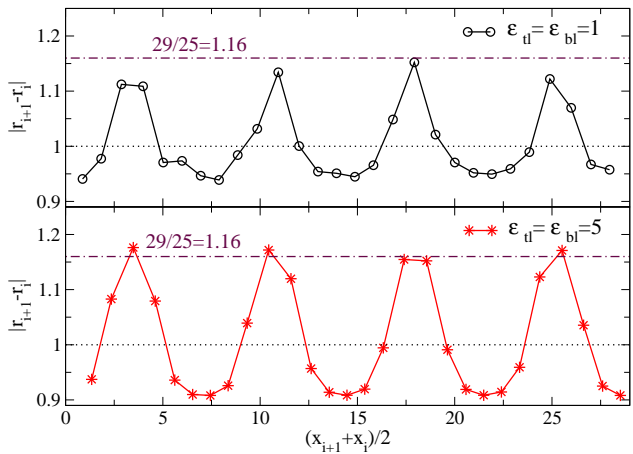}
} \caption{\label{snapshot:fig} (color online).
Snapshot of the dynamically pinned state of the 2D lubricant,
showing
the bond lengths as a function of the bond
horizontal position along the slider.
Shorter bonds $\approx a_{\rm b}$ 
identify
kink regions, while longer bonds
$\approx 1.16 a_{\rm b}$ belong to in-register regions.
Larger interaction with the substrates $\epsilon_{\rm tl} = \epsilon_{\rm
  bl} > 1$ favors the commensurate in-register region.
}
\end{figure}

The detailed dependence of the depinning transition on coverage is affected
by the relative ``softness'' of the lubricant.
Figure~\ref{eps_tp_bp:fig} shows that when the interaction energy
$\epsilon_{\rm tl} =\epsilon_{\rm bl}$ of the lubricant with the
substrates is increased to become
larger than the intra-lubricant interaction
$\epsilon_{\rm ll}$, as is the case
for a soft lubricant wetting
strongly the sliding surfaces, the quantized-plateau region
extends generally to larger $v_{\rm ext}$ and to a wider range of
coverages $\Theta$.
The reason for this is illustrated by the configuration snapshots of
Fig.~\ref{snapshot:fig}: a comparably stronger interaction with the
substrates $\epsilon_{\rm tl} =\epsilon_{\rm bl} > 1$ favors the
commensurate in-register lubricant regions,
shrinking the kink size. Localized kinks are more suitable to being
picked up and
dragged by the corrugations of
the top substrate.

\begin{figure}
\centerline{
\epsfig{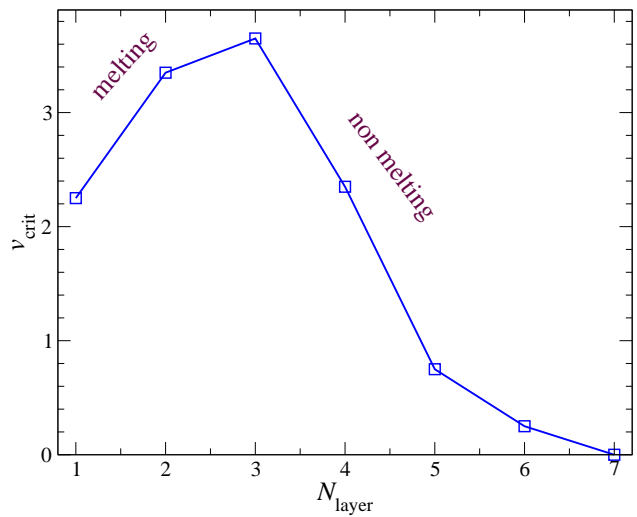}
} \caption{\label{multi:fig} (color online).
Critical depinning velocity $v_{\rm crit}$ as a function of the numbers
$N_{\rm layer}$ of lubricant layers.
All simulations are carried out in a condition that favors the
quantized-velocity sliding state: the model is composed by $4$, $29\cdot
N_{\rm layer}$ and $25$ atoms in the top lubricant and bottom layers
respectively (thus $\Theta=1$), with an applied load $F=25$ and $T=0.01$.
The data show an optimal dynamical pinning at $N_{\rm layer}=3$ and a
tendency for $v_{\rm crit}$ to drop considerably as the lubricant thickness
increases beyond that value.
For $N_{\rm layer}\geq 7$ no quantized plateau could be detected.
}
\end{figure}

Figure~\ref{multi:fig} shows how the lubricant thickness affects the
dynamically pinned state, i.e.\ $v_{\rm crit}$ as a function of the number
$N_{\rm layer}$ of lubricant layers in the fully commensurate $\Theta=1$
condition.
The data show quantized-velocity plateaus, displaying
a maximum extension for $N_{\rm layer}=3$, followed by a decaying
$v_{\rm crit}$ as the lubricant thickness increases beyond that
value.
For $N_{\rm layer}\geq 7 $ layers we could detect no velocity plateau, at
least within the $v_{\rm ext}$ range accessible to a practical numerical
integration of the equations of motion.
Vertical corrugations of the lubricant induced by the kinks
originating and propagating
from the bottom substrate have the effect of mediating the kink tendency to
pin to the top-layer corrugations, therefore favoring the observed perfect
velocity quantization.
For increasing $N_{\rm layer}$, these $z$-displacements decay rapidly, and
this mechanism becomes less effective to support the dynamically pinned
state.

Our calculations show that while the quantized plateau occurs when the
solitons -- misfit dislocations -- are driven gently through the solid
lubricant, the speed-induced transition to the $v_{\rm ext}>v_{\rm crit}$
dynamically unpinned state can involve the melting of the lubricant, due to
the large Joule heating induced by shearing.
By analyzing the detail of the depinning transition for
$\Theta=1$, we observe two different depinning behaviors to the
left and to the right, respectively, of the maximum shown in
Fig.~\ref{multi:fig}.
For $N_{\rm layer}\leq 3$, the depinning transition changes the solid
lubricant film into a disordered diffusive liquid, with a lower density,
(thus characterized by an increase in $\overline{r_{{\rm t}\,z}}$).
By contrast, a lubricant composed of four or more layers remains
essentially solid through the depinning transition.
This is the case because the Joule heating produced by shearing is removed
with sufficient efficiency by the thermostat so that the effective
lubricant temperature remains moderate.
The lubricant remains solid in particular when $v_{\rm crit}$ is small, as
is the case for thick lubricant layers, but also for $\Theta\neq 1$, or if
a larger damping $\eta\simeq 1$ is introduced in the Langevin dynamics.
This indicates that the quantitative details of the findings of the present
section depend on the assumed dissipation mechanism, and could therefore
vary between one or another specific experimental situation.
Nonetheless the qualitative result that under suitable conditions the
depinning could be associated to lubricant melting while under other
conditions the lubricant could remain solid through the transition is
expected to be real, and should be verifiable in real boundary-friction
experiments.

\section{Friction}
\label{friction:sec}

The ($x$-directed) kinetic friction force $F_k$ that must be applied to the
top slider to maintain its sliding motion balances instantaneously the
force that the lubricant exerts on the top substrate.
$F_k$ can be decomposed in several terms by analyzing the mechanical work
done by the individual force contributions.
The forces that the sliding top exerts on the lubricant [first term in
  Eq.~\eqref{Fj}] produce a total work $W_{\rm t\to l}$
whose mean
contribution to dissipation can be computed by evaluating the
change in lubricant kinetic energy over a time interval $\Delta
t$
\begin{eqnarray}
E_k(t\!&\!+\!&\!\!\Delta t)- E_k(t) =
\\\nonumber
&=&\sum_j\int_t^{t+\Delta t} \!\left[\vec F_j + f_{{\rm damp}\ j}
+ \vec f_j(t)\right] \cdot \dot {\vec r}_j \, dt'
\\\label{nontrivial}
&=&\sum_j \int_t^{t+\Delta t} \! \vec F_j \cdot \dot {\vec r}_j
\,dt' -\frac{4\eta}m \, \overline{E_k} \, \Delta t +
\\\nonumber
&+& \eta N_{\rm l} \int_t^{t+\Delta t}\!\! \vec v_{\rm cm}\cdot
\dot {\vec r}_{\rm t} \,dt' + \sum_j \int_t^{t+\Delta t} \vec f_j
\cdot \dot {\vec r}_j \,dt'.
\end{eqnarray}
The first term in Eq.~\eqref{nontrivial} represents precisely the
contribution of the LJ top-lubricant interactions to $W_{{\rm t\to
l}}$, plus bounded terms which do fluctuate, contributing a limited
[order $O(\Delta t)^0$] potential-energy change in the limit of a
long time integration.
The second term is proportional to the kinetic energy averaged over the
$\Delta t$ time interval.
The last term in \eqref{nontrivial} describes the correlation of
the random forces and the lubricant velocities.
By careful integration over the trajectory produced by the Langevin
dynamics, averaging over the distribution of the random variables $f_j$
(indicated by angular brackets), and recalling Eq.~\eqref{2fd}, this term
is evaluated as
\begin{eqnarray}\label{derivation1}
&&\langle\int_t^{t+\Delta t} \vec f_j(t') \cdot \dot {\vec r}_j(t')
 \,dt'\rangle =
\\
&&=\frac 1m
\int_t^{t+\Delta t} \!\!\!dt' \int_t^{t'}\!\!\!dt''
\,\langle\vec f_j(t') \cdot \vec f_j(t'')\rangle
\\\label{linewith2}
&&=\frac 1m \int_t^{t+\Delta t} \!\!\!dt' \int_t^{t'}\!\!\!dt''
\,4 \eta k_{\rm B} T \, \delta(t'-t'') \times 2
\\\label{derivation4}
&&=\frac {8 \eta k_{\rm B} T}m \int_t^{t+\Delta t} \!\!\!dt'\,
\frac 12 =\frac {4 \eta k_{\rm B} T}m \,\Delta t \,.
\end{eqnarray}
The factor 2 in line \eqref{linewith2} comes from summing $\hat x$ and
$\hat z$ components, and the factor $\frac 12$ in the integral of line
\eqref{derivation4} is the result of integrating the Dirac delta at the
integration boundary.
We observe that, like the conservative potential energy terms, the
lubricant kinetic energy deviation $E_k(t+\Delta t)- E_k(t)$ is of order
$O(\Delta t)^0$ over a long time integration.
By combining Eq.~\eqref{nontrivial} with the result of
Eq.~(\ref{derivation4}), in this large-$\Delta t$ limit the
average dissipated work becomes
\begin{eqnarray}\label{lubricantdiss}
W_{\rm t\to l} &=& \frac{4 \eta}m\, \Big[ \overline{E_{k}} -
N_{\rm l} k_{\rm B}T - \frac{N_{\rm l}}4 m
\,\overline{\vec v_{\rm cm}\cdot \dot {\vec r}_{\rm t}}
\\
&&+\frac{N_{\rm l}}4 m \,\overline{(\dot{{\vec r}}_{\rm t}-\vec
v_{\rm cm})\cdot \dot{{\vec r}}_{\rm t}}
 \Big]
\,\Delta t + O(\Delta t)^0
\,.
\end{eqnarray}
The fourth and last term in square brackets is added to account for the
work done by the top substrate directly against the dissipation forces of
Eq.~\eqref{fextratop}.
We omit the explicit indication of the averaging over the Langevin noise,
which becomes irrelevant in the limit of a long averaging time $\Delta t$.

This total dissipated energy is conveniently written in a form
where we measure the particles velocities relative to the
instantaneous center-mass velocity $\vec v_{\rm cm}$.
Indeed, by using $\dot {\vec r}_i = (\dot {\vec r}_i-\vec v_{\rm
cm}) + \vec v_ {\rm cm}$ and omitting corrections of order
$O(\Delta t^0)$, we can rewrite the dissipated energy as
\begin{eqnarray}
W_{\rm t\to l}
&=&
  \frac {4 \eta}{m} \left(
\overline{E_{k\,\rm cm}}- N_{\rm l} k_{\rm B}T
\right)\Delta t
\\\nonumber
&&+
 \eta N_{\rm l} \left[
 \overline{ v_{\rm cm}^2}
  + \overline{(\dot {\vec r}_{\rm t} - \vec v_{\rm cm})^2 }
\,\right]\Delta t
\,,
\end{eqnarray}
where terms linear in $(\dot {\vec r}_i-\vec v_{\rm cm})$ vanish due to the
definition of $\vec v_{\rm cm}$, and $E_{k\,\rm cm}$ is the kinetic energy in
the instantaneous CM frame:
\begin{equation}
E_{k\,\rm cm} = \frac{m}{2} \sum_i (\dot{\vec r}_i-\vec v_{\rm cm})^2
\,.
\end{equation}

\begin{figure}
\centerline{
\epsfig{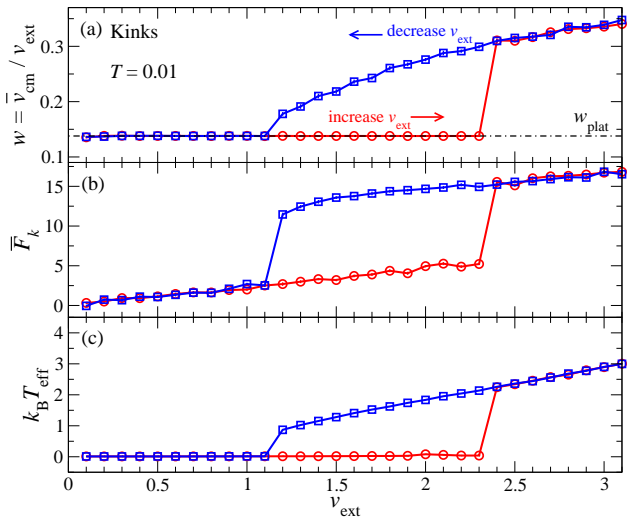}
}
\caption{\label{hyst-diss-kink:fig} (color online).
Results of the same model as in Fig.~\ref{vextmonotheta1:fig} ($F=25$,
$T=0.01$).
As a function of the top-layer velocity $v_{\rm ext}$ adiabatically
increased (circles) or decreased (squares) the three panels report:
(a) the average velocity ratio $w=\overline{ v_{{\rm cm}\,x}} /v_{\rm
  ext}$, compared to the plateau value $w_{\rm plat} =\frac{4}{29}\simeq
0.138$, Eq.~(\ref{wplat});
(b) the average friction force experienced by the top layer;
(c) the effective lubricant temperature, computed using the average kinetic
energy in the frame of reference of the instantaneous lubricant center of
mass, Eq.~\eqref{Teff}.
}
\end{figure}

\begin{figure}
\centerline{
\epsfig{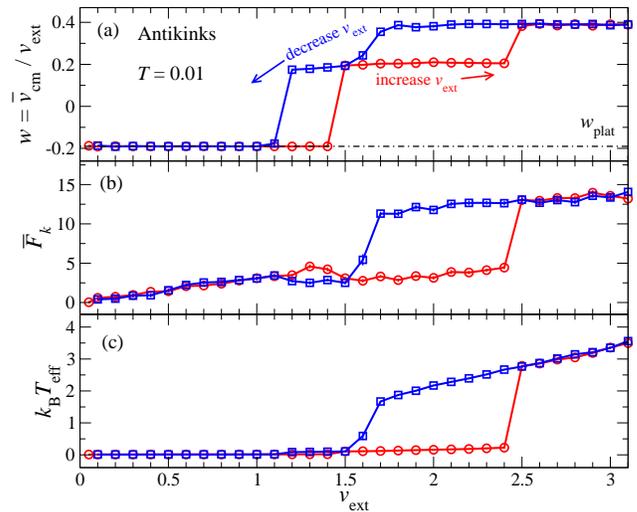}
}
\caption{\label{hyst-diss-anti:fig} (color online).
Results of a model composed by $4$, $21$ and $25$ atoms in the top,
lubricant and bottom chains, $\lambda_{\rm b}=21/25=0.84$, which according
to Eq.~(\ref{wplat}), produces perfectly quantized dynamics at a {\em
  negative} $w_{\rm plat}=-\frac{4}{21} \simeq -0.190$, a dot-dashed line
in panel a:
this backward lubricant motion is caused by forward-dragged anti-kinks.
The other simulation parameters are $F=25$, $T=0.01$.
As a function of the top-layer velocity $v_{\rm ext}$ adiabatically
increased (circles) or decreased (squares) the three panels report:
(a) the average velocity ratio $w=\overline{ v_{{\rm cm}\,x}} /v_{\rm ext}$;
(b) the average friction force experienced by the top layer;
(c) the effective lubricant temperature, Eq.~\eqref{Teff}.
}
\end{figure}

The work that the top layer does on the lubricant is supplied by the
kinetic friction force $F_k$ [the top-substrate vertical motion only yields
  a work of order $O(\Delta t^0)$], and equals $F_k\, v_{\rm ext}\,\Delta
t$.
This relation provides an instructive decomposition of the average kinetic
friction force:
\begin{eqnarray}\label{totalfriction}
\overline{ F_k} &=&
\frac 1{v_{\rm ext}\,\Delta t} W_{\rm t\to l}\\\nonumber
&=&
  \frac {4 \eta}{m\,v_{\rm ext}} \left(
\overline{ E_{k\,\rm cm}}- N_{\rm l} k_{\rm B}T \right)
\\\nonumber &&+ \frac{\eta N_{\rm l}}{v_{\rm ext}}  \left[
 \overline{ v_{\rm cm}^2}
  + \overline{ (\dot {\vec r}_{\rm t} - \vec v_{\rm cm})^2 }
\,\right]
.
\end{eqnarray}
Equation \eqref{totalfriction} expresses the kinetic friction force
$\overline{F_k}$ as the sum of two contributions.
The first and most important term is the contribution of the {\em
  fluctuations} of the lubricant-particle velocities in the instantaneous
CM frame, reduced by the thermostat-set value.
Even at $T=0$, this term would vanish in the event that the lubricant moved
as a rigid whole.
The term in the final line can be interpreted as the friction force that
dissipative phenomena with the bottom and the top respectively would
produce, again even if the lubricant was rigid.
This trivial term related to the overall CM motion yields a minimum total
friction force whenever $\vec v_{\rm cm}=\frac 12 \, \dot {\vec r}_{\rm
  t}$, which corresponds to $v_{{\rm cm}\,x}=\frac 12 \,v_{\rm ext}$ for
the main horizontal component.
In fact, this trivial contribution originates the tendency of $w= \overline{
  v_{{\rm cm}\,x}}/v_{\rm ext}$ to abandon the plateau and reach 0.5, as in
Figs.~\ref{vextmonotheta1:fig}, and \ref{hyst-diss-kink:fig},
\ref{hyst-diss-anti:fig} in the large sliding speed limit where this
trivial term dominates.
Equation~\eqref{totalfriction} can be reformulated in terms of an effective
temperature of the steady state:
\begin{eqnarray}\label{Teff}
k_{\rm B} T_{\rm eff}
&=&
\frac 1 {N_{\rm l}}\, \overline{ E_{k\,\rm cm} }
\\\nonumber
&=&
k_{\rm B} T
+ \frac m4 \, \left[
\frac {v_{\rm ext}} {\eta\, N_{\rm l}} \overline{ F_k}
- \overline{ v_{\rm cm}^2}
- \overline{ (\dot {\vec r}_{\rm t} - \vec v_{\rm cm})^2}
\,\right]
,
\end{eqnarray}
The formulation \eqref{Teff}, clarifies the contribution of friction to
Joule heating, and shows explicitly the irrelevant role of the trivial
dissipation terms in this respect.

Our MD simulations allow us to calculate instantaneously the ($x$-directed) kinetic friction by considering the longitudinal force component that the lubricant exerts on the top driven substrate.
By time averaging, the $\overline{F_k}$ value obtained coincides precisely with that evaluated through Eq.~\eqref{totalfriction}.

A direct numerical evaluation of the friction force is shown in panels (b) of Figs.~\ref{hyst-diss-kink:fig} and \ref{hyst-diss-anti:fig} for the kink and the anti-kink geometries, respectively, depicting typical scans where $v_{\rm ext}$ is cycled adiabatically up and down in small steps. Like in the $T=0$ simulations of the 1D sliding
model,\cite{Manini07PRE,Vanossi07Hyst,Vanossi07PRL} and as is well known for the static pinning/depinning of the standard FK
model,\cite{Braun97,Vanossi04,VanossiJPCM} the present 2D model displays a clear hysteretic behavior of the velocity-plateau state (panels (a)), provided that thermal fluctuations are much smaller than the energy barrier hindering the full exploration of phase space for any affordable simulation time.
A clear correlation of $\overline{F_k}$ with the effective lubricant
temperature $k_{\rm B} T_{\rm eff} \propto \overline{ E_{k\,\rm cm}}$
(panels Fig.~\ref{hyst-diss-kink:fig}c and \ref{hyst-diss-anti:fig}c) is evident, but the trivial contribution is also seen to play a quantitative role.
The bistability region with a clear hysteretic loop allows us to gauge the tribological effect of the quantized plateau state.
In the kink geometry, Fig.~\ref{hyst-diss-kink:fig} makes it apparent that friction is significantly less in the pinned quantized state than in the depinned state, all other parameters being equal.
Figure~\ref{hyst-diss-anti:fig} shows a similar tribological behavior, but
with the occurrence of an additional intermediate approximately quantized
velocity plateau (the 1D model displayed similar intermediate plateaus
\cite{Vanossi06}).
The difference in friction between the primary and the secondary plateaus
is a small one, with slightly less friction in the latter state.
Comparison of the curves in panels \ref{hyst-diss-anti:fig}b and
\ref{hyst-diss-anti:fig}c demonstrates that this effect is entirely due to
the trivial friction term.
This term is very large in the primary plateau because of the great
difference of $w$ from the optimal symmetrical drift $w=0.5$; in passing to
the secondary plateau, the trivial friction term decreases quite
substantially, but the kinetic fluctuation contribution to friction
increases by a similar amount.

\section{Intermittent dynamics at the plateau state}
\label{softnessintermittent:sec}

So far, we assumed some fixed externally imposed speed for
the top slider.
In a standard tribological simulation, the slider is
instead
generally pulled
through a spring of given stiffness whose other end is moved at constant
velocity.\cite{VanossiJPCM}
The spring can be viewed as a way to mimic not only the experimental
driving device, but also the elasticity of the sliding substrates.
 From an experimental point of view, frictional forces may display a typical
sawtooth dependence at sufficiently low driving external velocities, the
hallmark of the intermittent stick-slip dynamics.
The details of this low-driving behavior of course depend sensitively on the
mechanical properties of the device that applies the stress.

\begin{figure}
\centerline{\epsfig{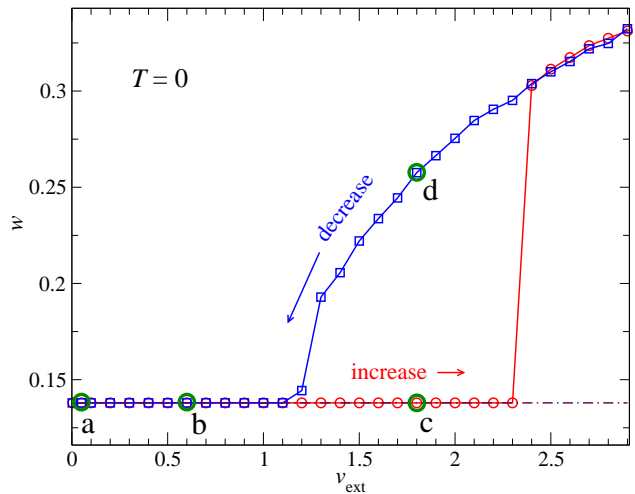} }
\caption{\label{Hyster_Tzero:fig} (color online).
The average velocity ratio $w=\overline{v_{{\rm cm}\,x}}/v_{\rm ext}$ as a
function of the adiabatically increased (circles) or decreased (squares)
top-layer velocity $v_{\rm ext}$.
The model parameters are the same as in Fig.~\ref{hyst-diss-kink:fig}, but
with no thermal effects ($T=0$).
}
\end{figure}

\begin{figure}
\centerline{\epsfig{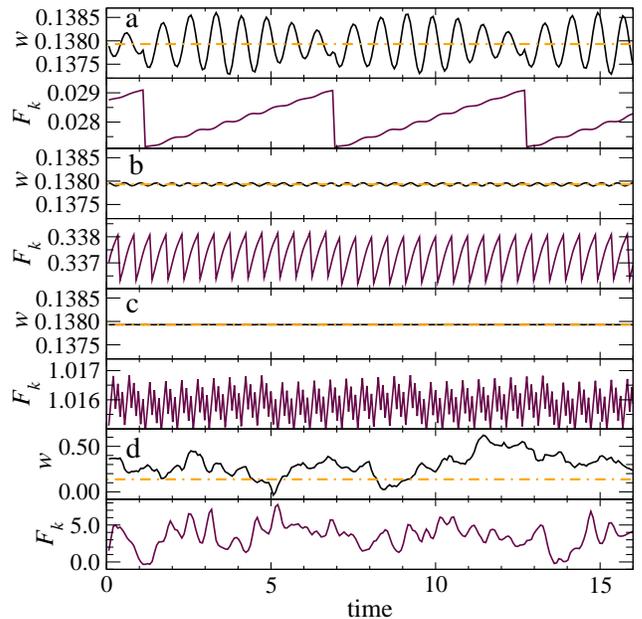}
} \caption{\label{stick_slip:fig} (color online).
The time evolution of the average velocity ratio $w$ and the corresponding
kinetic friction $F_k$ for the four dynamical states (a), (b), (c), and (d)
marked in Fig.~\ref{Hyster_Tzero:fig}.
The first three panels, referring to quantized sliding states, display a
typical intermittent stick-slip dynamics with small amplitude fluctuations;
the last panel, not corresponding to a plateau state, exhibits chaotic
large jumps in both $w$ and $F_k$.
the dot-dashed lines highlight the quantized-plateau value $w_{\rm plat}$.
}
\end{figure}


Beside constant speed, we also explored the model with the spring pulling method.
We obtained the quantized sliding state, also with the spring, and the simulations showed differences with the constant-speed method only in the fine detail.
For this reason we omit here to display and analyze those results in detail.

In fact, even within the rigid-drive model, for sufficiently low sliding
speed $v_{\rm ext}$ and for extremely small thermal fluctuations (i.e.,
$T\simeq 0$), we did observe a stick-slip intermittent dynamical regime in the
total force applied to the sliding top substrate in order to keep its
sliding speed constant.
Figure~\ref{Hyster_Tzero:fig} depicts a typical adiabatic up-down scan of
$v_{\rm ext}$, similar to Fig.~\ref{hyst-diss-kink:fig}a but at $T=0$.
To get insight into the different dynamical regimes outside and inside the
bistability region, Fig.~\ref{stick_slip:fig} analyzes in detail the four
points marked in Fig.~\ref{Hyster_Tzero:fig}, three of which belong to the
plateau state and one outside of it.
Each of the four panels of Fig.~\ref{stick_slip:fig} compares the time
evolution of the lubricant CM velocity to the corresponding instantaneous
kinetic friction force $F_k$.

For the first two plateau points (a) and (b) at low $v_{\rm ext}$, the
typical sawtooth time dependency of $F_k$ is clearly visible.
In these regimes of motion, the rescaled lubricant CM velocity $w = v_{{\rm
    cm}\,x}/v_{\rm ext}$ performs tiny periodic oscillations around the
quantized plateau value and $F_k$ exhibits an intermittent, almost regular,
stick-slip pattern.
This oscillatory phenomenology is understood as the kink array being
dragged at full velocity $v_{\rm ext}$ across the associated periodic
Peierls-Nabarro potential.\cite{Braunbook}
Indeed, the $F_k$ oscillation frequency matches the washboard frequency of
the kinks $v_{\rm ext}/a_{\rm b}$ multiplied by their number $N_{\rm
  kink}=N_{\rm t}=4$ in these $\Theta=1$ simulations, due to the four
equally-spaced kinks crossing sequentially the corresponding
Peierls-Nabarro barriers.
A much less regular pattern would arise in a  $\Theta\neq1$ geometry.
Due to the rather extended nature of solitons and smaller amplitude of the
Peierls-Nabarro barrier with respect to the full atomic corrugation, the
quantized plateau stick-slip regime is associated to much lower dissipation
than regular stick-slip dynamics as seen in dry friction AFM
experiments.\cite{Verhoeven04,Maier08}
By further increasing the external driving speed to $v_{\rm ext}=1.8$,
panel (c), the kinetic friction becomes a little more ``erratic'', with an
extra modulation of the sawtooth shape, resembling a sort of inverted
stick-slip.
Panel (d) displays the dynamics for the same $v_{\rm ext}=1.8$, but
obtained while cycling $v_{\rm ext}$ down: this depinned state shows large
fluctuations both in $w$ and $F_k$, giving rise to significant tribological
dissipation, as expressed by the first term of Eq.~\eqref{totalfriction}.

\section{Discussion and conclusions}
\label{conclusions:sec}

The present simulation work is meant to characterize tribologically the quantized sliding state discovered in a 1D sliding model \cite{Vanossi06} and later observed in a substantially less idealized 2D model.\cite{Castelli08Lyon}
The perfectly quantized plateau is demonstrated here to extend over broad parameter ranges of the model, being robust against the effects of thermal fluctuations, quenched disorder in the confining substrates, the presence of confined multiple (up to 6) lubricant layers, and over a wide interval of loading forces.
When temperature becomes comparable to the lubricant melting point, the plateau state tends to deteriorate and eventually disappears, but the geometrically determined velocity ratio $w_{\rm plat}$, acting as an attractor, leaves anyway a trace in the ensuing ``noisy'' dynamics.

The velocity plateau, as a function of $v_{\rm ext}$, ends at a critical velocity $v_{\rm crit}$, and for $v_{\rm ext}>v_{\rm crit}$ the lubricant tends to accelerate toward a speed $0.5\,v_{\rm ext}$.
This result, dictated by the symmetric choice of the Langevin thermostat dissipation, may change in real systems where heat dissipation occurs through generally asymmetric sliders.

Our calculations show that while the quantized plateau occurs when the solitons are driven gently through the solid lubricant, the speed-induced transition at $v_{\rm ext}>v_{\rm crit}$ can involve the melting of the lubricant, due to the large Joule heating induced by large velocity shearing.The depinning value $v_{\rm crit}$ is linked to the rate of commensuration $\Theta$ of kinks to the upper slider periodicity: a particularly robust plateau signaled by a local maximum of $v_{\rm crit}$ is located at well-commensurate $\Theta$ values, especially $\Theta=1$. Besides, the lubricant softness, setting the width of the propagating solitonic structures, is found to play a major role in promoting in-registry contact regions beneficial to this quantized sliding. Our typical dynamical depinning speed $v_{\rm crit}$, is of the order of a few model units (corresponding to sliding speed values ranging from tens to thousands m/s for realistic choices of model parameters), is very large compared to typical sliding velocities investigated in experiments. This suggests that in practice sliding at a dynamically quantized velocity is likely to be extremely robust. In real experimental systems, deterioration of the quantized sliding state will most probably associated to mechanisms such as disorder, boundary effects, or unfavorable lubricant-substrate incommensurate geometries, rather than to excessive driving speed.

By cycling $v_{\rm ext}$ in underdamped regime, the layer sliding velocity exhibits a hysteretic loop around $v_{\rm crit}$, like in the 1D model.\cite{Vanossi07PRL} The bistability region allows us to gauge the tribological effect of the quantized plateau state. So, in the framework of Langevin dynamics, by evaluating the force instantaneously exerted on the top plate, we find that this quantized sliding represents a dynamical ``pinned'' state, characterized by significantly low values of the kinetic friction. A characteristic backward lubricant motion produced by the presence of ``anti-kinks'', has also been observed in this context. This peculiar backward motion is likely to represent the most curious evidence of the quantized plateau state when -- as we hope -- it will be investigated experimentally.

On the theoretical side, the role of substrate deformability and defects in the lubricant structure, together with a realistic three dimensional description, with force fields representative of a concrete lubricated configurations, will certainly require further investigations.

\section*{Acknowledgments}
This work was supported by CNR, as a part of the European Science
Foundation EUROCORES Programme FANAS. R.C.\ and A.V.\ acknowledge gratefully the financial support by the Regional Laboratory InterMech - NetLab ``Surfaces \& Coatings for Advanced Mechanics and Nanomechanics'' (SUP\&RMAN), and of the European Commissions NEST Pathfinder program TRIGS under Contract No.\ NEST-2005--PATHCOM-043386.



\end{document}